\documentstyle[12pt]{article}
\begin{document}
\begin{titlepage}
\hoffset = -1truecm
\centering
\title{Particle with internal dynamical asymmetry: chaotic
self-propulsion and turning}
\author{S. Denisov\\
{\it Laboratory for Turbulence Research, Institute for Single
Crystals},\\ {\it Lenin av. 60, Kharkov 61001, Ukraine}\\ e-mail:
denisov@isc.kharkov.com}

\date{14th July 2000}

\maketitle

\begin{abstract}
We consider model of a complex particle that consists of a rigid
shell and a nucleus with spatial asymmetric interaction. The
particle's dynamics with the  nucleus driven by a periodic
excitation is considered. It is shown that unidirectional
self-propulsed particle motion arises in the absence of spatial
and temporary asymmetry of external potentials and influences.
Transport modes are the  general case of complex particle dynamics
in the presence of nonlinear friction or periodic external
potential. The changes of average transport velocity and direction
of transport are determined by qualitative changes of internal
dynamics regimes: local attractor bifurcations in the internal
phase space of the complex particle. Finally,  microbiological
relevance of the proposed  model is briefly discussed.\\{\bf PACS
number}: 05.40.+a,  05.60.+k
\end{abstract}
\end{titlepage}

\voffset = -1.5truecm \hoffset = -1truecm
 Over the past years, the appearance and
intensive development of some areas of nonlinear dynamics and
statistical physics have been inspired by  several microbiological
phenomena [1-4]. First, it is the physical principles and
functional mechanisms of molecular motors,  nano- and microscale
engines, which effectively  convert chemical energy to mechanical
work [1]. Second, it is physics of motility, namely the
unidirectional transport of microbiological objects [2]. Both
these research lines are closely related to each other, and  the
biological  motor efficiency  can  be defined  in the terms of
average  transport  velocity[3].

One of the most interesting mechanisms of the directed transport
of microobjects is the so-called ratchet-effect, related to
particle motion  in periodic potential with spatial asymmetry [4].
As a rule, the problem of the ratchet mechanism is considered
within the frame of stochastic approach, when the unidirectional
motion of overdamped Brownian particle in asymmetrical potential
takes places due to unbiased nonequilibrium fluctuations in  the
absence of any macroscopic forces and gradients [4-5]. However,
recently a significant interest has been paid to the deterministic
approach to this problem,  taking into account the finiteness of
inertial terms [6-7]. A deterministic description allows to expand
dynamic modes spectrum, including regular and chaotic ones [8],
and  to study the role of interaction mechanisms in greater
detail. Within the framework of this approach,  similar to the
deterministic diffusion phenomenon [9], low-dimensional dynamic
chaos can effectively play the role of   thermodynamic
fluctuations. In [7], Mateos studied the event of the
deterministic motion of a particle in smooth ratchet potential
under  the action of a symmetrical periodic time-dependent force.
He showed that a wide spectrum of dynamic regimes, with
bifurcation transitions between regular and chaotic regimes,
arises when  periodic force amplitude is varied. Moreover,  he
found, that  current reversals are  related with bifurcations from
chaotic to periodic regimes. In a recent paper [10], Flach,
Yevtushenko and Zolotaryuk have studied a  more general case of a
directed current appearance in  deterministic nonlinear systems.
It has been shown that the key factor for a transport regime
appearance  is the existence of spatial ("spatial ratchet") and/or
temporary ("temporal ratchet") asymmetry of influences upon a
particle.

Thus, one of the necessary triggers for directed transport
appearance is the existence of asymmetry in the system. The basic
idea of the present paper is that this asymmetry can be "built"
into the internal degrees of freedom of the complex particle and,
in this way, the source of motion can be located inside of the
object itself.   At that, external potentials and the influences
can be completely symmetrical  [11]. Such a particle is an active
self-propulsed walker, which can control the direction and
velocity of its own motion. Further we will formulate a simple
one-dimensional model of a complex particle and we will consider
its dynamics in:  (i) a medium with nonlinear friction (that is
true for real biological media nearly always [12]) and (ii) in a
linear medium under external periodic symmetrical potential. We
will analyse spectrum of dynamic regimes that arises under
changing parameters of the system, and also the relations between
the peculiarities of a complex particle internal dynamics and its
directed transport characteristics. Finally, we will shortly
discuss the biological motivation of the proposed model.

Let us consider a one-dimensional system, consisting of a rigid
spherical shell with a mass $M$, whose position is determined by
the coordinate of its centre $X$, and an internal particle
("nucleus") with a mass $m$  and the coordinate $x$, attached to
the internal walls of the shell (see inset in Fig1.). The elastic
interaction between the shell and the nucleus is governed by an
asymmetric nonlinear potential $U(x-X)$, so that the force arising
due to the nucleus displacement from the centre of the shell
depends not only on the absolute displacement value $\Delta x=x-X$
, but also on its sign (Fig.1). The minimum of the potential is
located at the origin ($U^{'}(x)=0$), which corresponds to the
equilibrium nucleus position in the centre of the shell. Let us
also presume that the nucleus is driven by a periodic symmetrical
force $f(t)=a cos(\omega t)$ which is defined, for example, by
mechanochemical coupling to a periodic biochemical reaction, such
as  ATP hydrolisis [13]. Then, generally, with an external
potential and nonlinear friction present, the  complex particle
dynamics, after trivial mass normalisation, is defined by the pair
of equations:
\begin{eqnarray}
\ddot{x}=-K_{1}(\dot{x}-\dot{X})-{\partial U(x-X) \over \partial
x} +a cos(\omega t),\\ \ddot{X}=-K_{2}(\dot{X})+ \mu {\partial
U(x-X) \over \partial x} +W^{'}(X) \label{1}
\end{eqnarray}
\\
where $K_{i}(\nu)$ are  nonlinear friction forces,
$\mu=\frac{m}{M}$ is the mass factor,  $a, \omega$  are driving
force amplitude and frequency respectively.
        The potential of interaction is given by (Fig.1):
\begin{eqnarray}
U(x)=\alpha (x-x_{0})^2+\beta (x-x_{0})^4-\gamma (x-x_{0})^3
\label{3}
\end{eqnarray}
\\
where $\alpha , \beta, \gamma$  are the parameters of interaction,
$x_{0}$ is an appropriate shift, such  that a minimum of potential
(3) is located at the origin. Nonlinear friction forces are given
by $K_{i}=s_{i}\nu +q_{i}\nu^{3}$. The parameters $\mu =2,
\omega =1,  \alpha =\beta =1,  \gamma =2.5,  x_{0}\approx 1.553$
are fixed throughout this paper.

The complete phase space of the system (1-2) is five-dimensional
$(x, X, \dot x, \dot X, t)$, the dynamic equations are nonlinear
and contain dissipative terms. So, in the phase space of system
(1-2) both periodic (limiting cycles) and chaotic (strange
attractors) attractors can exist [8].  On the other hand, the
complex particle structure induce, in the natural way, the
splitting of the complete phase space into two subspaces, the
external one, accessible to  an external observer - $(X, \dot X,
t)$, and the internal one, corresponding to an  observer located
within the shell - $(x-X, \dot x - \dot X, t)$. The first subspace
corresponds to nonlocal transport of the complex particle as a
whole, the second one corresponds to local dynamics, which is
determined by the internal interaction. The most interesting
question is  how the correlation between these  two scales of  the
particle's complete dynamic is realised. While zero-mean driving
symmetrical periodic force $f(t)$ does not contain a constant
component, the appearance  of the directed current can only be a
consequence of the asymmetry of the internal interaction.

Let us consider now the motion of a particle in a nonlinear medium
in the absence of an external potential ($W(x)\equiv 0, s_{1}=0.2,
s_{2}=0.5, q_{1}=10^{-2}, q_{2}=5\cdot 10^{-3}$). For the analysis
of the particle's dynamics in the external phase space ($X, \dot
X, t$) we used stroboscopic Poincare section [8] with the period
equal to the driving force period, $T=2 \pi /\omega$. The
bifurcation diagram for $V=\dot X$ in a limited range of the
parameter $a$ is shown in Fig.2a  [14]. At first, the standard
scenario of period-doubling route to chaos is realised ($a\in
[5.6, 6.0347]$), and then the bifurcation connected with the
internal crisis of a chaotic attractor ($a_{1} \approx 6.2117$)
and the opposite tangent bifurcation connected with the birth of a
steady period-three cycle take place ($a_{2} \approx 6.96441$)
[8]. In Fig.2b we show the current $J=\frac{1}{NM} \sum_{j=1}^{M}
\int_{t_{0}}^{t_{0}+N} V_{j}(t) dt$, $(t_{0}=50, N=10^{4}, M=50)$,
as the function of parameter $a$ in the same range of values. As
one can see, current reversal takes place exactly at the tangent
bifurcation point $a_{2}$. As viewed from the internal phase space
$(x-X, \dot x-\dot X, t)$ it corresponds to transition from the
chaotic attractor (Fig.3a) to the periodic one (Fig.3b). Besides,
a smaller  jump of the current value takes place at the internal
crisis bifurcation point $a_{1}$, that corresponds to a sudden
expansion of chaotic attractor [15]. It can be shown that by a
certain variation of the system parameters it is possible to
achieve the current reversal taking place exactly at this
bifurcation point. The investigation of the model's dynamics in
other ranges of parameter $a$ showed, that any abrupt change of
the current $J$ is connected with sudden changes of the internal
chaotic attractor - {\it crisis} (as interior crisis at $a_{1}$),
or {\it subduction} (as tangent bifurcation at $a_{2}$) [15].
Here, current reversal with the a greater probability occurs at
bifurcations of the second type, that is tangent bifurcations (as
in this case the jump of a current value is much greater).

Let us consider now the motion of a complex particle under
external periodic potential $W(x)=A cos(X)$.  To separate
nonlinear friction mechanisms, let us presume that  the medium is
linear $(q_{i}\equiv 0)$. We found  that in this case, as well as
in the one considered above, there also exist transport modes.
However, unlike the free motion case, in this case the transport
modes take place within certain parameter ranges. It is connected
with the presence of external potential, which  adds to the system
some additional temporal and spatial characteristic scales, while
some coherence between the parameters of external potential and
internal force and interaction is needed for the appearance of
transport. We found a very interesting effect which is connected
with these coherence mechanism, the effect of sharp change
(current "switching") (Fig.4b). In the   parameter space this
effect looks like full overbarrier reflection, when the particle
flying in the ballistic mode above potential $(V_{m}=\int_{0}^{T}
V(t) dt = 1, a \in [5.1, 5.206])$ changes its flight direction to
the opposite $(V_{m}=\int_{0}^{T} V(t) dt = -1, a \in [5.363,
5.57])$ at a small variation of the driving force amplitude. The
corresponding bifurcation diagram for $V$ is shown in Fig.4a. As
one can see,  the current switching, as it is in  the nonlinear
friction case, is connected with tangent bifurcation  at
$a_{c}\approx 5.3633$ leading to  the birth of the limiting cycle
from the chaotic attractor. The   period-one limiting cycle of the
particle internal phase space $C_{1}$ corresponds to the ballistic
transport  in the positive direction in the region, while another
period-one cycle, $C_{2}$, corresponds to the transport in the
negative direction (see inset in Fig.4a).

The existence of these limiting cycles, $C_{1}$ and $C_{2}$,
allows to distinguish between two characteristic spatial scales
and to separate the solution  in the external phase space into two
parts $X(t)=X_{s}(t)+\xi (t)$, where $X_{s}(t)$ is a slow part and
$\xi (t)$ is a small fast part. After that, using perturbation
theory ideology [16], for the slow variable we can construct a
"quasi" zero-order approximation:
\begin{eqnarray}
\ddot X_{s}=-s_{2} X_{s} + W^{'}(X_{s})+S(t) \label{4}
\end{eqnarray}
where
\begin{eqnarray}
S(t)=G(x(t)-X(t))={\partial U(x(t)-X(t)) \over \partial x(t)}
\label{5}
\end{eqnarray}
is a periodic force acting on the shell that corresponds to the
exact solution of the system \nolinebreak (1-2). This forces
$S(t)$ is zero-mean, $\int_{0}^{T} S(t)dt = 0$, with pronounced
temporal asymmetry (Fig.5a and Fig.6a). Thus, the equation (4)
describes the motion of a {\it simple} particle under a periodic
potential, driving by a spatially homogeneous {\it asymmetric}
periodic force [10]. For the numerical integration of the equation
(4) we used as excitation force the first six harmonics of the
complete solution, $S^{*}(t)=\sum_{k=1}^{6} S_{k} cos(k\omega t+
\phi_{k})$. The obtained results demonstrate very close agreement
with the solutions of the entire system (1-2). In $C_{1}$ - case
in the in phase space of system (4) there exists the only one
global attractor that corresponds to the particle ballistic
transport  in the positive direction (Fig.5a), and in $C_{2}$ -
case there is the only one global attractor that corresponds to
the particle ballistic transport  in the negative direction
(Fig.5b). In the case of chaotic transport (e. g., $a\in [2.208,
5.363]$) approximation (4) is not valid, because due to spectrum
continuity there are no distinguished scales in the system.

It is necessary to note, that in each of the considered cases
(free motion in nonlinear medium and motion  under external
periodic potential), there is only one global attractor (limit
cycle or chaotic attractor) in the phase space of system (1-2) for
the  given value of parameter $a$. In those cases   averaging over
ensemble is equivalent to averaging over attractor invariant
measure, and so we can talk about the directed motion  of an
individual self-propulsed complex particle rather than  of
current. The obtained current $J$ dependencies  can be represented
as dependencies of the average drift velocity $V_{m}$ of the
particle, and we also can talk about turning instead of current
reversal.

In summary, we have shown, that the internal dynamic asymmetry,
"dynamical chirality ", can be a source of the directed motion of
a complex particle. Transport modes are not exceptional, but
occupy  certain parameter ranges  (it is a general case for the
motion in  a nonlinear medium). The direction and velocity of the
motion are determined by  the particle internal dynamics
properties, and their changes are connected with the internal
attractor bifurcations. It is interesting to note, that a similar
situation will take place, if we consider a particle with
symmetrical nonlinear internal interaction $U(x)$  and a temporal
asymmetrical driving periodic force  $f(t)$ [17]. Such a force may
be stimulated by a limiting cycle of some biochemical reaction
[18], which is conformally coupled with the internal particle
structure. Here, the force temporal asymmetry  will be a general
case, while symmetry is possible as an exceptional case in the
space of reaction parameters.

The assumption of absolute rigidity of the shell is not a strict
requirement. The validity of such assumption corresponds to the
validity of  the inequality $\theta = \omega_{s} / \omega \ll 1$,
where $\omega$ is the driving force frequency and $\omega_{s}$ is
the characteristic frequency of large-scale deformation vibrations
of the shell. Moreover, the viscosity of the external and internal
medium  also impedes the excitation of such oscillations. The
energy dissipation by the small shell vibrations can be taken into
account within nonlinear friction terms.

The considered mechanism can shed light upon physics of some
non-canonical forms of  microbiological motility. For example,
swimming Cyanobacteria, having the shape of almost exact spheroid
and  having no  external appendages perform directed motion in a
liquid without any observable shape changes [19]. The mechanism of
internal dynamic asymmetry can be additional to the hydrodynamic
mechanism accounting for the transport of Cyanobacteria through to
small-scale periodic modulations of the shell shape [20]. In the
case of more complex cells, the internal structure of the particle
can be a concentrated image of the cytoskeletal structure [13].
Cytoskeletal polymers all have complex nonlinear viscoelastic
characteristics [21]. Moreover, interactions between cytoskeletal
polymer systems in living cell may result in composite material
with unique nontrivial dynamic properties [22]. Furthermore, the
proposed model provides for  dynamic modelling of such collective
phenomena,  as microorganisms interaction and self-organisation at
the level of microscopic equations of the motion of a single
biological particle. The microscopic approach to these phenomena
that consider a  microorganism as an active object in active
environment [23], makes it possible to effectively use  the ideas
about  interaction and synchronization of internal attractors  of
individual microobjects [24].

For valuable discussions, we are indebted  to  Prof. V. V.
Yanovsky,  Prof.  O. V. Usatenko and Dr. Hab. S. Flach.

\newpage

\begin{center}
{\bf FIGURE CAPTIONS}
\end{center}

1. The interaction potential $U(x)$  between shell and nucleus.
Inset shows the complex particle structure.

2. (a) Bifurcation diagram for $V$ as a function of $a$ and (b)
the current $J$ as a function $a$  for $\mu=2,  \omega=1,
\alpha=\beta=1,  \gamma=2.5,  W(X)\equiv0,  s_{1}=0.2,  s_{2}=0.5,
q_{1}=10^{-2},  q_{2}=5\cdot 10^{-3}$.

3. Projections of complex particle internal phase space
attractors: (a) chaotic attractor for $a=6.96$  and (b)  the
period-three limit cycle just below tangent bifurcation at
$a\approx6.9644$.

4. (a) Bifurcation diagram for $V$ as a function of $a$  and (b)
the current $J$ as a function  $a$ for $A=3,  s_{1}=0.5,
s_{2}=0.8,  q_{1}=q_{2}\equiv 0$. Inset shows (a) internal phase
space period-one limit cycles: $C_{1}$ ($a=5.18$, positive
current) and $C_{2}$ ($a=5.41$, negative current) and (b)
corresponding particle trajectories.

5. (a) The shape of periodic zero-mean asymmetric force $S(t)$,
corresponding to limit cycle $C_{1}$ ($a=5.18$, positive current):
solid curve - exact solution for system (1 - 2), dotted curve -
first six harmonics sum; (b) corresponding limit cycle in the
reduced external phase space ($V_{m}=1$): solid curve for full
system (1), dotted curve - for approximation (4) with the first
six harmonics as driving force.

6. Same as in Fig.5 for cycle $C_{2}$ ($a=5.41$, negative
current).

\begin{thebibliography}{99}
\bibitem{1} M. O. Magnasco, Phys. Rev. Lett. {\bf 72}, 2656 (1994);
T. Duke, S. Leibler, Biophys. J. {\bf 70}, 689 (1996); R. D.
Astumian, Science {\bf 276}, 917 (1997); F. Julicher, A. Ajdari,
and J. Prost, Rev. Mod. Phys. {\bf 69}, 1269 (1997).
\bibitem{2} P. Hanggi  and R. Bartussek, in Nonlinear Physics of
Complex Systems, Lecture Notes in Physics, Vol. 476, ed. By J.
Parisi, S. C. Muller, and W. Zimmermann (Springer, Berlin, 1996),
pp. 294-308.
\bibitem{3}R. Yasuda, H. Noji, K. Kinosita, and M. Yoshida,
Cell {\bf 93}, 1117 (1998); I. Derenyi, M. Bier, and R. D.
Astumian, Phys. Rev. Lett. {\bf 83}, 903 (1999).
\bibitem{4} M. O. Magnasco, Phys. Rev.
Lett. 71, 1477 (1993); J. Maddox, Nature {\bf 369}, 181 (1994); J.
Rousselet, L. Salome, A. Ajdari, and J. Prost, Nature {\bf 370},
446 (1994); C. Doering, W. Horsthemke, and J. Riordan, Phys. Rev.
Lett. {\bf 72}, 2984 (1994); I. Derenyi and T. Vicsek, Physica A
{\bf 249}, 397 (1998).
\bibitem{5} Various models of overdamped Brownian ratchets are
as follows: C. S. Peskin, G. M. Odell, and G. F. Oster, Biophys.
J. {\bf 65}, 316 (1993); J.Luczka, R. Bartussek, and P.  ,
Europhys. Lett. {\bf 31}, 431 (1995); H.-X. Zhou, Y. - d. Chen,
Phys. Rev. Lett. {\bf 77}, 194 (1996); F. Julicher  et al., Phys.
Rev. Lett. {\bf 78}, 4510 (1997); A. Vilfan, E. Frey, F. Scwalbe,
Eur. Phys. J. B. {\bf 3}, 535 (1998).
\bibitem{6} P. Jung, J. G. Kissner, and P. Hanggi,  Phys. Rev. Lett. {\bf 76},
3436 (1996).
\bibitem{7} J. L. Mateos, Phys. Rev. Lett. {\bf 84}, 258 (2000).
\bibitem{8}E. Ott, {\it Chaos in
Dynamical Systems} (Cambridge University Press, Cambridge, 1993).
\bibitem{9}S. Grossmann and H. Fujisaka, Phys. Rev. A 26, 1179 (1982); P.
Gaspard, J. Stat. Phys. {\bf 68}, 673 (1992).
\bibitem{10}S. Flach, O.
Yevtushenko, and Y. Zolotaryuk, Phys. Rev. Lett. {\bf 84}, 2358
(2000).
\bibitem{11} We consider symmetry here, following [10], as existence, in
the case of potential $W(X)$, some appropriate argument shift
$\chi$, such that $f(X+\chi)=-f(-X+\chi)$,  ($f(X)=-dW(X)/dX)$);
as existence, in the case of homogenous periodic force $E(t)$,
some appropriate time shift $\tau$, such that
$E(t+\tau)=E(-t+\tau)$ or $E(t)=-E(t+\tau)$.
\bibitem{12}H. A.
Barner, J. F. Hutton, and K. Walters, {\it An Introduction to
Rheology} (Elsevier, Amsterdam, 1989).
\bibitem{13}L. Darnell, {\it Molecular Cell
Biology} (W. H. Freeman, San Francisco, 1990).
\bibitem{14}This range of $a$ is
chosen for presentation. Transport modes take place for any
non-zero value of driven force amplitude.
\bibitem{15}C. Grebogi and E. Ott, Physica D {\bf 7}, 181
(1983).
\bibitem{16}A. H. Nayfeh, {\it Introduction to Perturbation
Techniques} (J. Wiley and Sons, NewYork, 1981).
\bibitem{17}S. Denisov, in preparation.
\bibitem{18}I. R. Epstein and J. A. Hojman, {\it An Introduction to Nonlinear
Chemical Dynamics: Oscillations, Waves, Patterns, and Chaos}
(Oxford University Press, Oxford, 1998).
\bibitem{19}J. B. Waterbury, J.
M. Willey, D. G. Franks, F. W. Valois, and S. W. Watson, Science
{\bf 230}, 74 (1985).
\bibitem{20}H. A. Stone and A. D. T. Samuel, Phys. Rev.
Lett. {\bf 77}, 4102 (1996).
\bibitem{21}P.  Janmey, Curr. Opin. Cell. Biol. {\bf 3},
4 (1991);
\bibitem{22}F. Mackintosh, J.,  and P. Janmey, Phys. Rev. Lett.
{\bf 75}, 4425 (1995).
\bibitem{23}W.
Ebeling, F. Scweitzer, and B. Tilch, BioSystems {\bf 49}, 17
(1999); B. Tilch, F. Schweitzer, and W. Ebeling, Physica A {\bf
273}, 294 (1994); F. Schweitzer, B. Tilch, and W. Ebeling, Eur. J.
Phys. B (in press).
\bibitem{24} L.Pecora, T. Carroll, G. Johnson, and D. Mar, Chaos {\bf 7},
520 (1997).
\end{thebibliography}
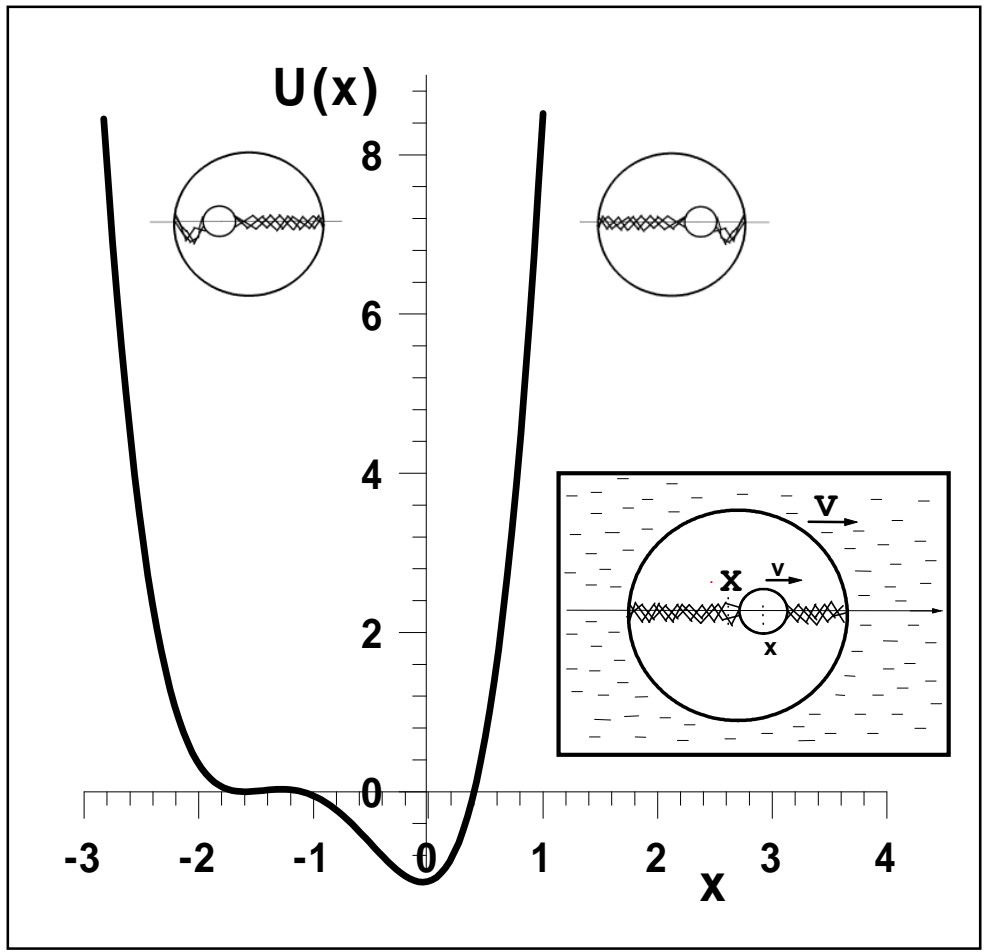
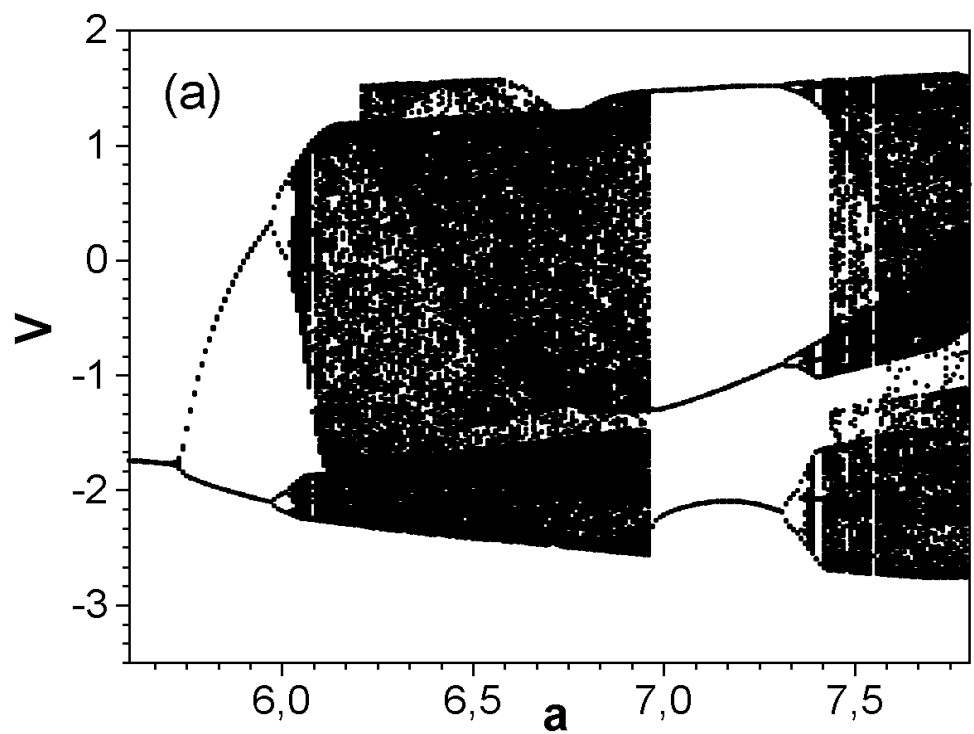
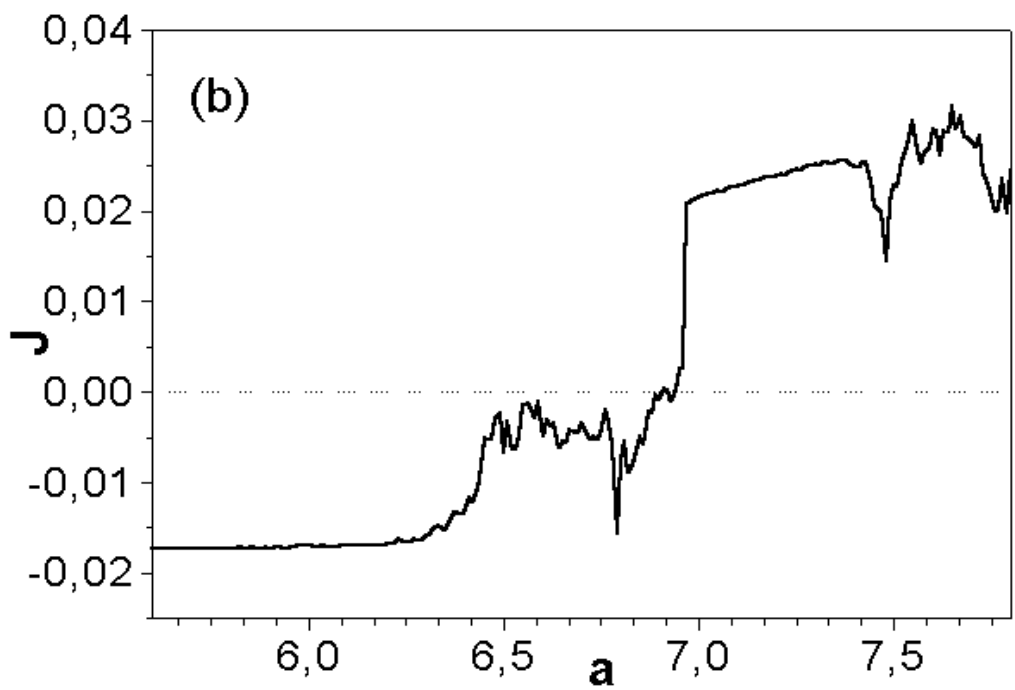
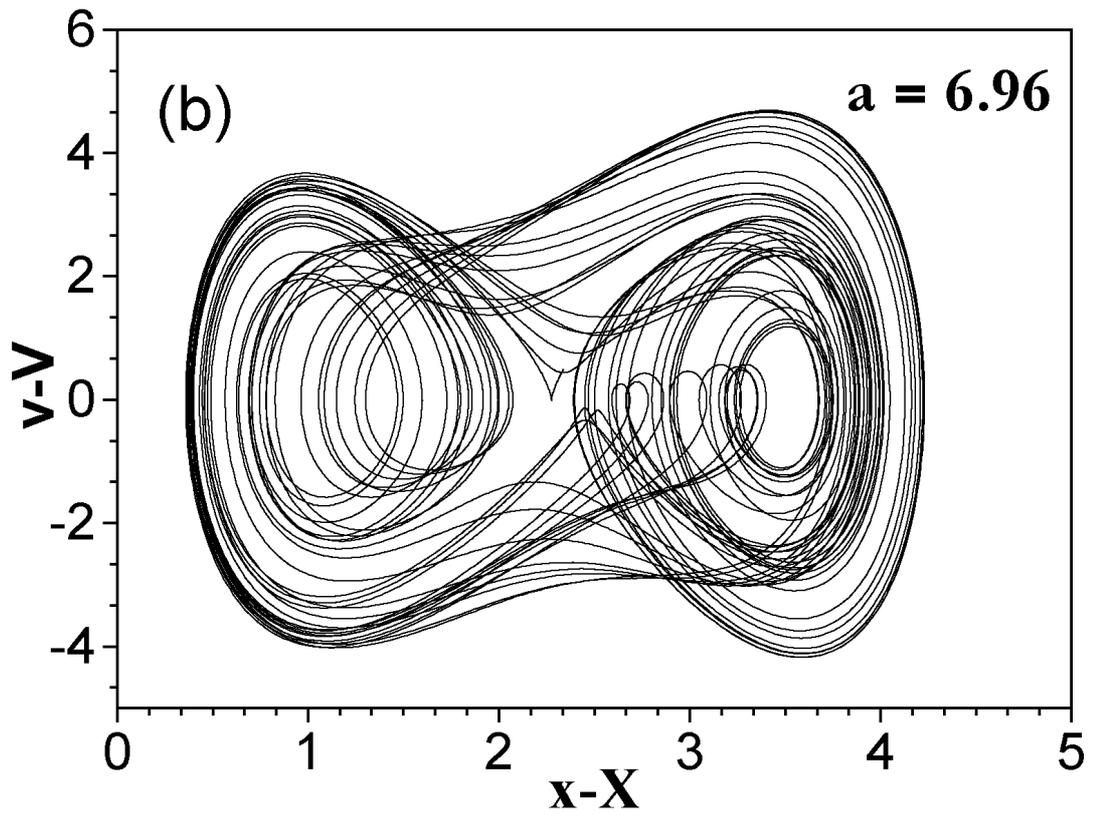
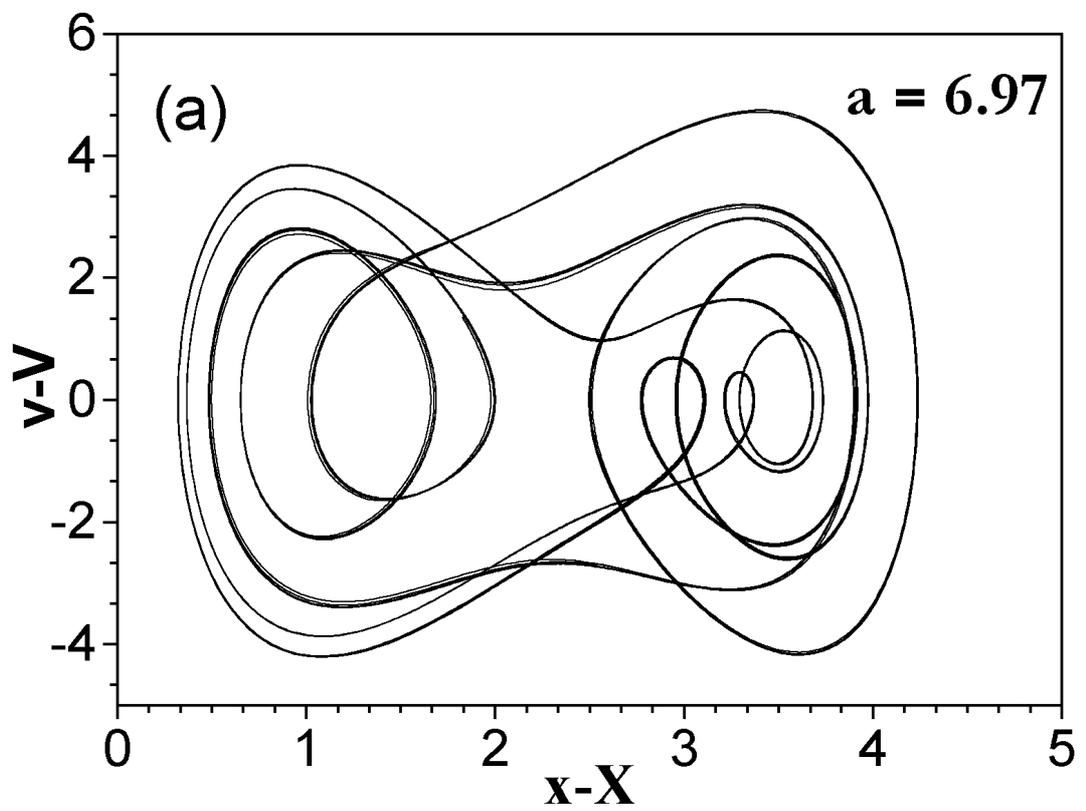
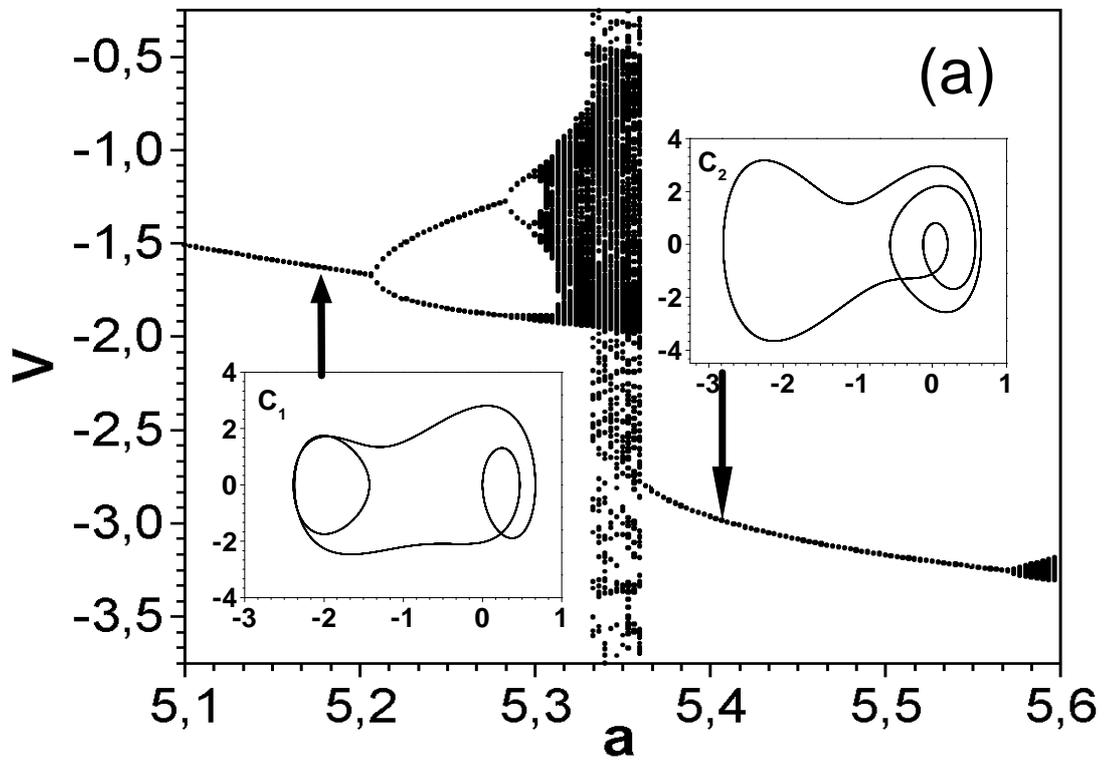
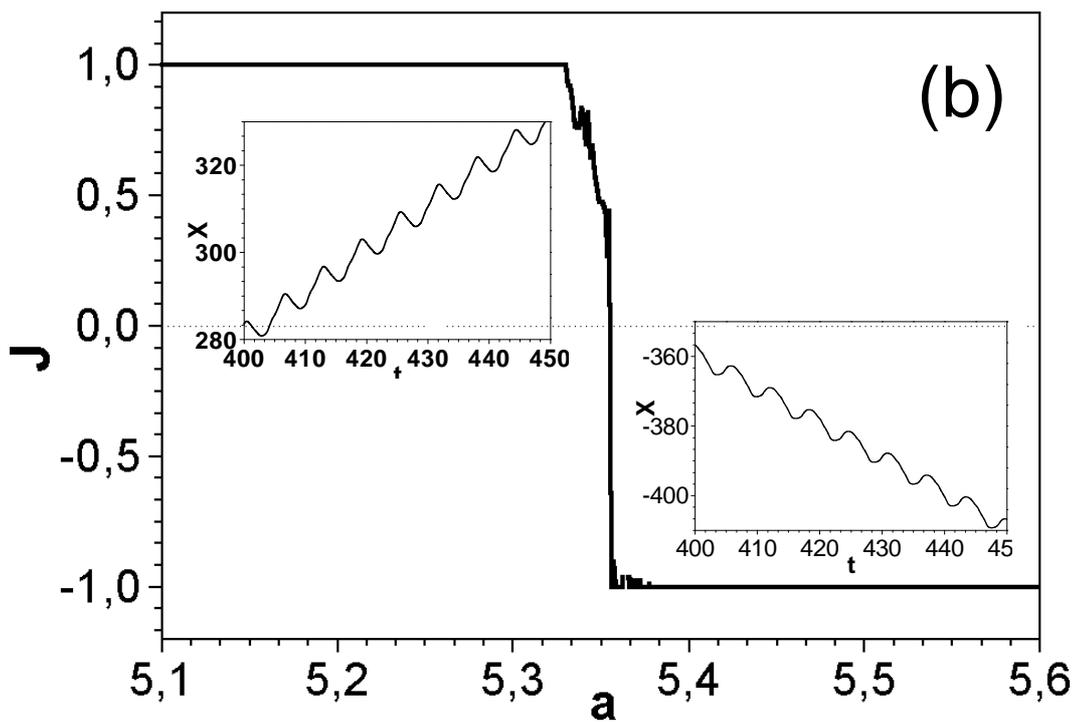
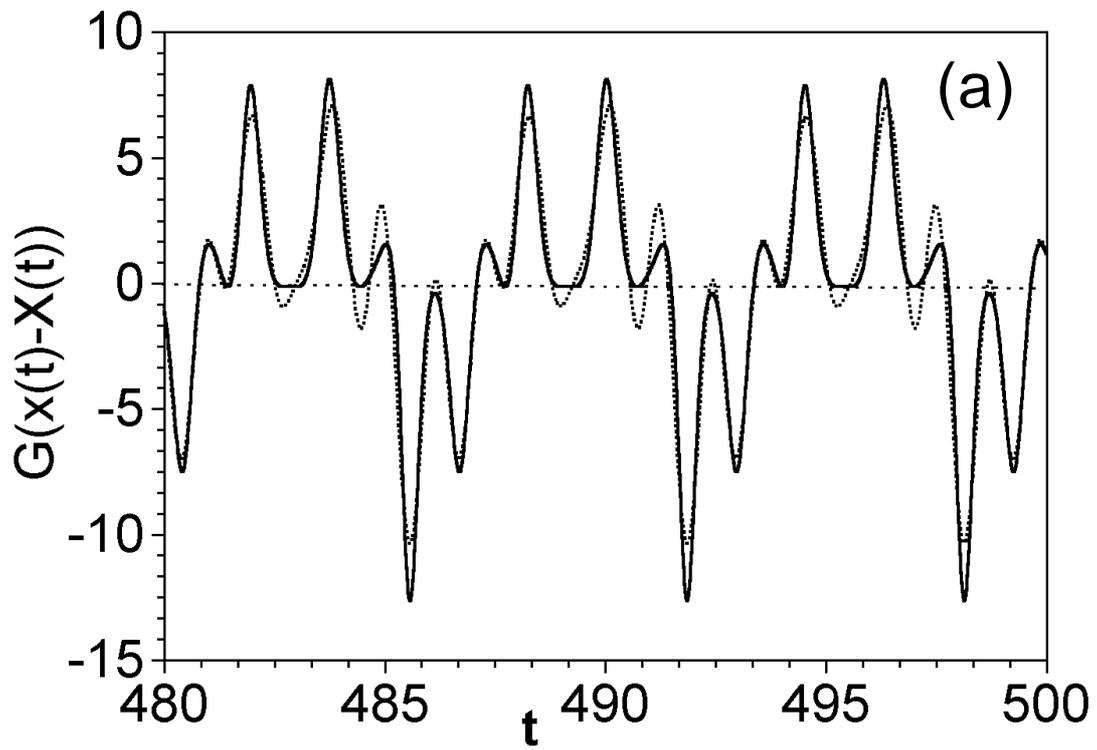
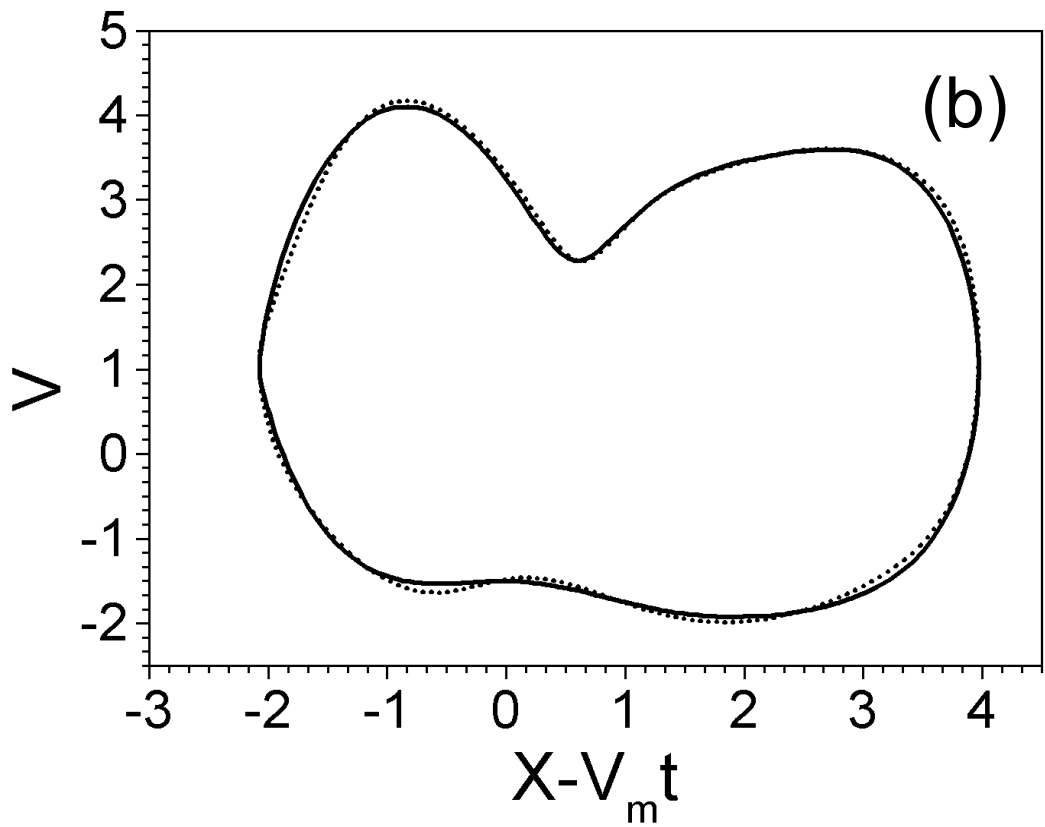
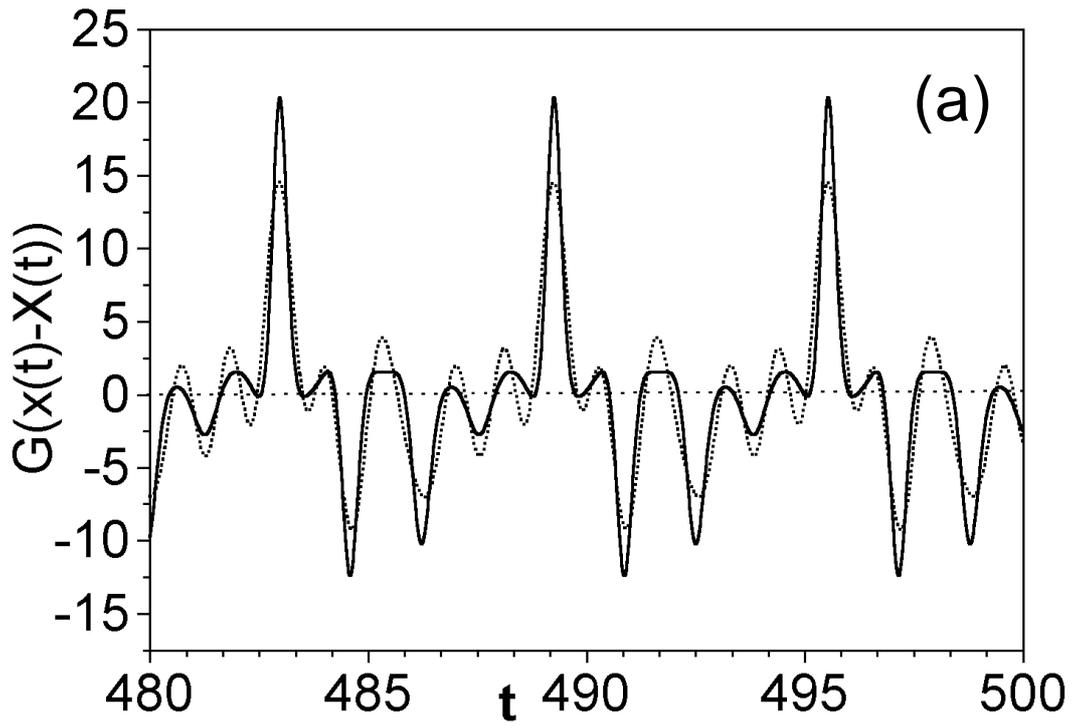
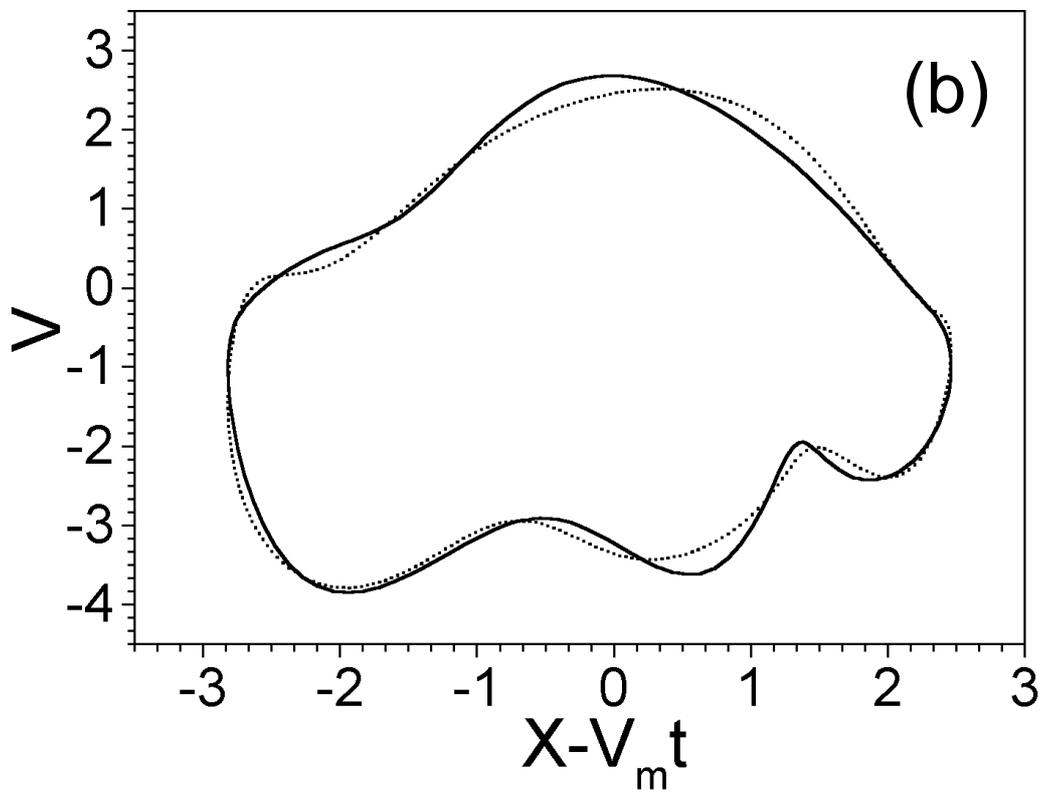
\end{document}